\documentclass[%
 reprint,
 amsmath,amssymb,
 aps,
]{revtex4-2}

\usepackage{graphicx}% Include figure files
\usepackage{dcolumn}% Align table columns on decimal point
\usepackage{bm}% bold math
\usepackage{xcolor}
\newcommand{\pd}{\partial}
\newcommand{\mc}{\mathcal}

\begin{document}

\preprint{APS/123-QED}

\title{Robustly encoding certainty in a metastable neural circuit model}% Force line breaks with \\

\author{Heather L Cihak}
\email{Heather.Cihak@colorado.edu}
\author{Zachary P Kilpatrick}%
\email{zpkilpat@colorado.edu}
\affiliation{Department of Applied Mathematics, University of Colorado, Boulder, Colorado 80309, USA}

\date{\today}

\begin{abstract}
    Localized persistent neural activity can encode delayed estimates of continuous variables. Common experiments require that subjects store and report the feature value (e.g., orientation) of a particular cue (e.g., oriented bar on a screen) after a delay. Visualizing recorded activity of neurons along their feature tuning reveals activity \textit{bumps} whose centers wander stochastically, degrading the estimate over time. 
    Bump position therefore represents the remembered estimate. Recent work suggests bump amplitude may represent estimate certainty reflecting a probabilistic population code for a Bayesian posterior. Idealized models of this type are fragile due to the fine tuning common to constructed continuum attractors in dynamical systems. Here we propose an alternative metastable model for robustly supporting multiple bump amplitudes by extending neural circuit models to include \textit{quantized} nonlinearities.
    Asymptotic projections of circuit activity produce low-dimensional evolution equations for the amplitude and position of bump solutions in response to external stimuli and noise perturbations. Analysis of reduced equations accurately characterizes phase variance and the dynamics of amplitude transitions between stable discrete values. More salient cues generate bumps of higher amplitude which wander less, consistent with the experimental finding that greater certainty correlates with more accurate memories.

\end{abstract}

\maketitle

\section{\label{sec:introduction}Introduction}
Working memory involves the essential ability to encode and store information for short periods of time~\cite{GoldmanRakic1995}. Since estimation errors can propagate through subsequent computations~\cite{raghubar2010working}, robust and flexible maintenance of information is key for daily tasks like decision making and planned movement~\cite{hulse2020mechanisms,romo1999neuronal,Funahashi89}. 
Delayed estimates of a continuous object feature value are encoded by persistent and spatially localized neural activity across multiple brain regions~\cite{Curtis06} sustained by feature-specific excitation and lateral inhibition~\cite{Constantinidis16,GoldmanRakic1995}. Estimate abnormalities can be indicators of neural dysfunctions arising in schizophrenia~\cite{stein2021towards}, autism~\cite{vattikuti2010computational}, and attention deficit hyperactivity disorder~\cite{klingberg2002increased}. Thus identifying mechanisms supporting working memory stability may guide diagnostics for predicting neuropsychopathologies~\cite{montague2012computational}. Biologically aligned computational models are useful for identifying how such disorders present and may also act as a testbed for intervention~\cite{wang2014computational,hodgson2019eye,klingberg2002increased}.

We focus here on extending neural circuit models of visuospatial working memory, building on decades of successful interaction between oculomotor delayed response experiments and physiologically inspired models~\cite{Funahashi89,Compte2000,Constantinidis16}. In the task, a subject must identify and remember the position of a briefly presented cue, and then indicate the remembered location after a few seconds. Neural recordings reveal that the centroid of neural activity bumps encodes the remembered location of the cue during the delay and response~\cite{Wimmer14}. Connections between pyramidal (excitatory) neurons maintain persistent activity during the delay and interneuron (inhibitory) populations help localize activity to those with similar feature tuning as the cue~\cite{GoldmanRakic1995}. Fluctuations in neural and synaptic activity cause the activity bump to wander diffusively, generating response error variance that scales roughly linearly with time~\cite{Wimmer14,Fuster73,White94,faisal2008noise}.

Subjects also appear to reliably encode confidence (or certainty) in their delayed estimates~\cite{honig2020humans}. Confidence reports align with accuracy, suggesting delayed estimates are represented probabilistically, possibly by the firing rate level of persistent activity in neurons encoding the estimate~\cite{bays_schneegans_ma_brady_2022} which has been observed to increase with training and higher working memory performance~\cite{tang2019working}. Peak neural activity during retention periods has been shown to increase with strength of evidence~\cite{romo1999neuronal,basso1997modulation}, consistent with Bayesian computation~\cite{beck2008probabilistic,kutschireiter2023bayesian,esnaola2022flexible}. Visual attention~\cite{mcadams1999effects}, stimulus presentation duration~\cite{groen2022temporal,badcock2008examining}, and cue contrast~\cite{groen2022temporal,mcadams1999effects} all can increase spike rates and corresponding estimates in neural circuits representing recalled sensory stimuli~\cite{cohen2014correspondence,mcgettigan2012speech,cazettes2016cue}. Overall, these experiments suggest increased (decreased) activity during delay periods generates higher (lower) certainty and more (less) accuracy in estimates~\cite{meyer2011stimulus,honig2020humans,basso1997modulation}.

\begin{figure*}
    \includegraphics[width=0.9\textwidth]{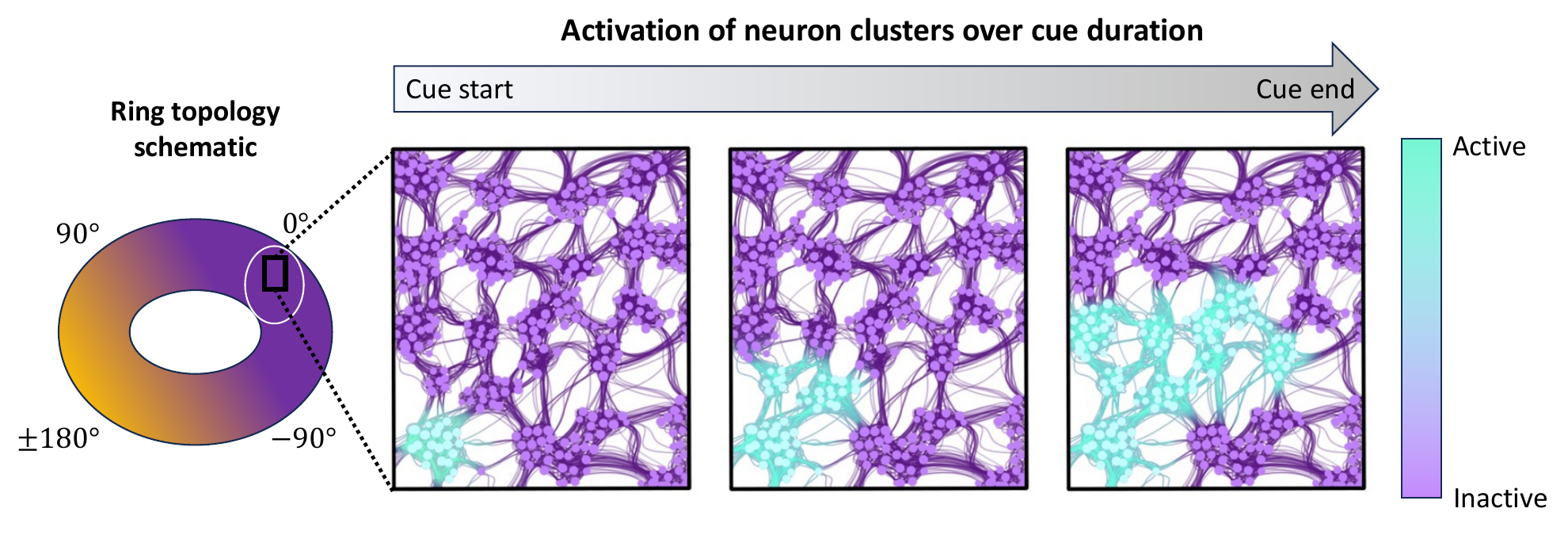}
    \caption{{\bf Schematic of local metastability in a ring attractor network.} Consistent with recording~\cite{la2019cortical,brinkman2022metastable} and modeling~\cite{bressloff2010metastable,litwin2012slow} studies of metastable neural circuits, stronger and/or longer stimuli successively and discretely recruit more active microclusters locally in a neural circuit. Such microclusters may emerge spontaneously in development due to interactions of self-sustained activity with neuronal migration and outgrowth~\cite{okujeni2019self,beaulieu2018enhanced}. Macroscale connectivity has ring topology akin to that inferred and observed in recordings of neural circuits encoding periodic continuum variables~\cite{zhang1996representation,Compte2000,hafting2005microstructure,kim2017ring}.}
    \label{fig1:cluster}
\end{figure*}

Our models relate neural activity amplitude and response errors along these lines. Building on physiologically inspired models~\cite{Wimmer14,Constantinidis16} and stochastic methods~\cite{Kilpatrick13WandBumpSIAD,cihak2022distinct} linking neural circuit activity to delayed estimates, we develop a theory of activity-based encoding of confidence and its impact on response accuracy. Larger amplitude bumps have steeper spatial profiles and wander less in response to fluctuations, better retaining estimates~\cite{Krishnan18}. The theory of bump attractors must be extended to consider how bump amplitude impacts estimate storage and readout~\cite{carroll2014encoding,kutschireiter2023bayesian}. Most circuit models support bumps of a single amplitude, generating a bistable amplitude space in which a stable quiescent state and stable wide bump are separated by an unstable narrow bump~\cite{Amari77,bressloff2019stochastic}. Bumps are either instantiated or not but cannot encode certainty in bump amplitude since they filter out richer cue information often represented in neural recordings~\cite{cohen2014correspondence}. Here we propose and analyze a mechanism for robust encoding of certainty in activity bumps with graded amplitude values which can be reached in response to variable stimulus features.

Metastability of firing rate states is a common phenomenon in the brain, observed across multiple timescales. Single neurons can occupy multiple possible discrete firing rates states without drive~\cite{goldman2003robust}, which could arise due to network level phenomena revealed in stochastically switching firing rate sequences~\cite{sasaki2007metastability,mazzucato2015dynamics}. Single neuron models with multiple bistable dendritic compartments exhibit robust and quantized firing rate sets, providing short term memory of transient inputs represented by quasi-continuous staircase-like firing rate functions~\cite{goldman2003robust}. Recent complementary work has suggested strong and sustained oscillatory input from intrinsic cell mechanisms or circuitry may work similarly~\cite{champion2023oscillatory}, producing phase-locked states with graded firing rate amplitudes. 
Alternatively, recorded macroscale neural population activity also exhibits multiple metastable states local attractor-like dynamics~\cite{sasaki2007metastability}. Furthermore, transitions between metastable state sequences observed in neural data are well captured by clustered population spiking models that globally exhibit discrete firing rate increases
% models with clustered microstructure also produce discretely graded population level firing rates when viewed at macroscale 
(See Fig.~\ref{fig1:cluster} and \cite{sasaki2007metastability,mazzucato2015dynamics,la2019cortical}). 

In contrast to metastable neural circuit models~\cite{bressloff2010metastable,litwin2012slow,thompson2016relating}, theory has also pursued finely tuned circuit models to support activity bumps with continuously graded amplitudes~\cite{carroll2014encoding,kutschireiter2023bayesian,esnaola2022flexible}. Fixing the gain of a piecewise linear firing rate function in spatially extended rate models generates activity bumps whose position and amplitude jointly lie on a planar continuum attractor~\cite{carroll2014encoding}: radial location encodes amplitude and angle encodes position. However, model perturbations destroy the line attractor~\cite{seung1996brain,lim2013balanced} and the bump amplitude wanders in response to noise. 
Such fragility is alleviated by breaking the symmetry of such continuum attractors, stabilizing a discrete set of attractors separated by saddles~\cite{brody2003basic,kilpatrick2013optimizing}; more aligned with the discrete and quasi-continuous firing rates sets examined in other studies~\cite{goldman2003robust, mazzucato2015dynamics, sasaki2007metastability, champion2023oscillatory}.

Thus, we introduce and analyze a neural circuit model supporting metastable dynamics akin to those observed and derived in a number of prior models~\cite{goldman2003robust,bressloff2010metastable,schaub2015emergence,brinkman2022metastable,champion2023oscillatory} and supported here with staircase shaped input-firing rate relationships. Metastability is conceived as arising from successive activation of neural microclusters with increasing cue salience (Fig.~\ref{fig1:cluster}). Stable activity bump solutions have multiple graded amplitudes allowing stimulus-dependent encoding of estimate certainty (Section \ref{sec:Model}), whose dynamics are characterized by reduced phase-amplitude equations (Section~\ref{sec:determ_C4}). Our model is more robust to perturbations than prior models with a continuum of amplitudes~\cite{carroll2014encoding} or an all-or-none (bistable) response~\cite{Amari77}. Bumps subjected to fluctuating inputs retain a roughly constant amplitude for long time intervals, and their amplitude dependent wandering dynamics can be determined from reduced equations (Section~\ref{sec:stochas_C4}).

\section{\label{sec:Model}Model equations}

\begin{figure*}[t!]
\includegraphics[width=0.85\textwidth]{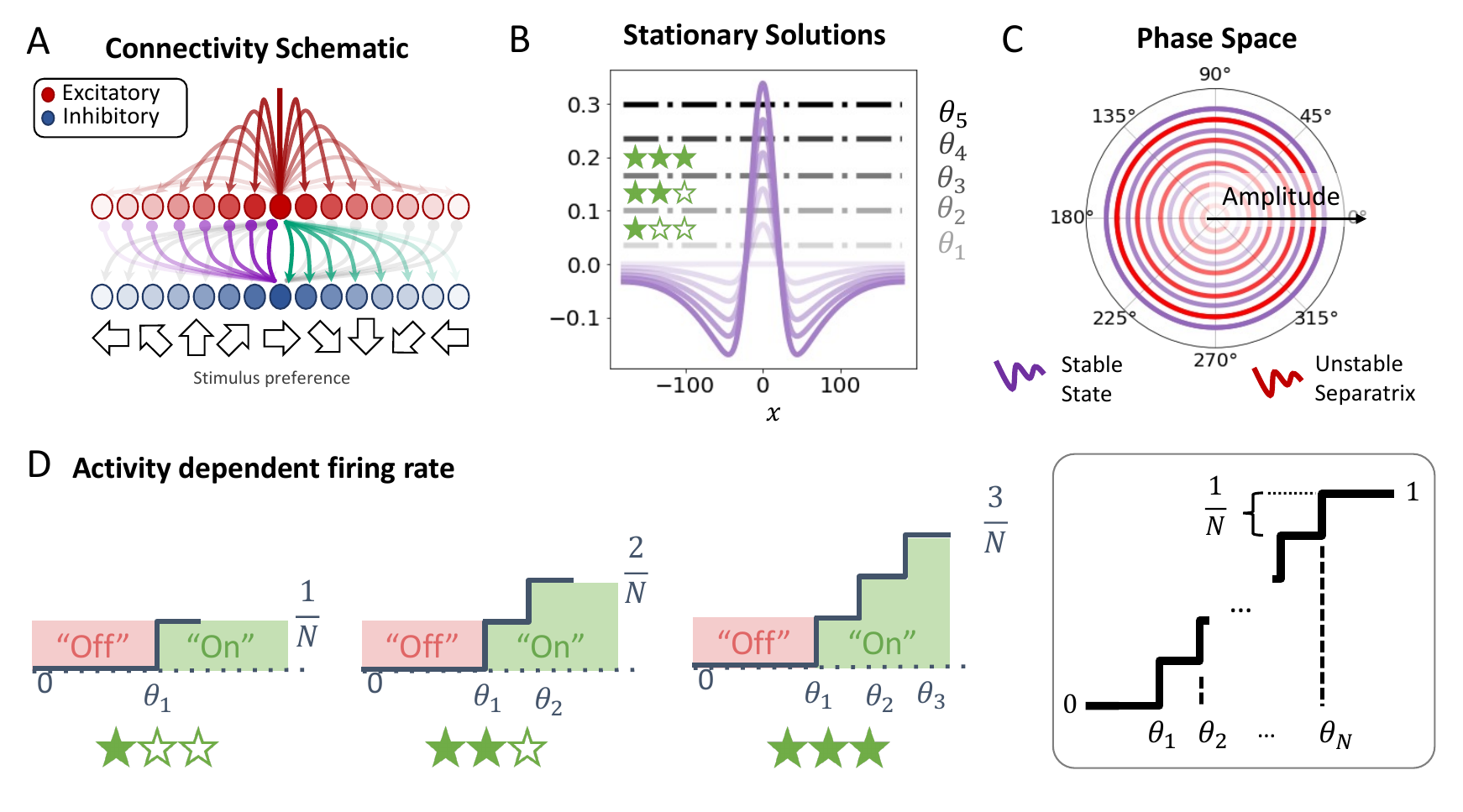}% Here is how to import EPS art
\caption{\textbf{Model structure and core dynamics.} {\bf A.}~Excitatory/inhibitory network connectivity depends on difference in stimulus angle preference. {\bf B.}~Broader and weaker inhibitory connectivity promotes stable and localized activity bumps which can exhibit multiple graded amplitude values due to stairstep firing rate nonlinearities (See panel {\bf D}). Bump activity $U(x)$ is plotted here as a function of angle $x$ (in degrees, not radians). {\bf C.}~Phase-amplitude space plots of bumps reveal concentric ring attractors separated by unstable ring repellers, stabilizing amplitude representations. {\bf D.} Increasing input successively engages higher firing rate states in the stairstep transfer function.}
\label{fig2:model}
\end{figure*}

\begin{table}[b!]
\vspace{-4mm}
{\footnotesize
  \caption{\textbf{Numerical \& model parameters for Eq.~(\ref{eq:model_C4})}}\label{tableparam_C4}
\begin{center}
\vspace{-2mm}
  \begin{tabular}{|c|c|c|} \hline
   \bf{Parameter} & \bf{Definition}& \bf{Value}\\ \hline
   $x$ & Domain & $[-180,180]$ degrees\\ \hline
   $dx$ & Spatial increment & $\frac{360}{n}$ where $n=2^{12}+1$\\ \hline
   $dt$ & Time step & $0.025$ \\ \hline
   $A_{e}$ &E strength & 1.5\\ \hline 
   $A_{i}$ & I strength & 0.5 \\ \hline
   $\kappa_{e}$ & E synaptic footprint & 20\\ \hline
   $\kappa_{i}$  & I synaptic footprint & 1\\ \hline
   $M$ & Fourier modes & 20 \\ \hline
   $N$ & Firing rate steps & 5 \\ \hline
   $\boldsymbol{\theta}$& Firing thresholds & [0.035,0.1,0.165,0.234,0.298] \\ \hline
   $A_{c}$ & Cue amplitude & 1\\ \hline
   $a_{c}$ & Cue radius & 0.02 radians or $\approx 1.15$ degrees \\ \hline
  \end{tabular}
  \vspace{-4mm}
\end{center}
}
\end{table}

Our network attractor model encodes an angle on the circle $\Delta \in [- \pi, \pi)$, a common requirement of memory and navigational tasks~\cite{zhang1996representation,Compte2000,kim2017ring} (See Table~\ref{tableparam_C4} for parameters). Excitatory and inhibitory neural populations are collapsed to a single neural field (integro-differential) equation organized with a ring topology. Effective input $u(x,t)$ to local clusters at time $t$ is indexed by angular preference along a continuum ($x \in [- \pi, \pi)$ for analysis, but sometimes converted to degrees $180x/\pi$ for plotting), and clusters with similar orientation preferences are strongly coupled by excitation while those with dissimilar preferences effectively inhibit each other~\cite{Amari77}. Evolution of network activity is described by the spatially-extended Langevin equation:
\begin{align}
    du(x,t)=&[-u(x,t)+\int_{-\pi}^{\pi}w(x-y)f(u(y,t))dy \nonumber \\&+I_{c}(x,t)]dt+\sqrt{\epsilon}dW(x,t). \label{eq:model_C4}
\end{align}
Recurrent connectivity targeting clusters $x$ from angular position $y$ is described by the effective synaptic kernel, $w(x-y)$ (Fig.~\ref{fig2:model}A), which is locally excitatory and laterally inhibitory. 

A single stationary bump solution is generated when considering a Heaviside step nonlinearity $f(u)=H(u-\theta)$ with $H(u-\theta) = 1$ if $u \geq \theta$ and 0 otherwise~\cite{Amari77,coombes2004evans,cihak2022distinct,cihak2024multiscale, Kilpatrick13WandBumpSIAD}. To incorporate certainty we examine a sequence of metastable amplitude states; generating stationary bump solutions (Fig.~\ref{fig2:model}B) of different amplitudes (Fig.~\ref{fig2:model}C) when we consider staircase firing rate functions with $N$ steps (Fig.~\ref{fig2:model}D)
\begin{align}
    f(u) = \frac{1}{N}\sum_{k=1}^{N}H(u-\theta_k). \label{stair}
\end{align}
% where the Heaviside step nonlinearity $H(u-\theta) = 1$ if $u \geq \theta$ and 0 otherwise. 
Each step on the staircase reflects successive cluster activations (similar to \cite{mazzucato2015dynamics}), prompting increased population level firing activity associated with distinct metastable states (Fig.~\ref{fig1:cluster}).
Appropriate choices of the thresholds $\boldsymbol{\theta} = [\theta_1, ..., \theta_N]$ provide for $N$ stable bump solutions (e.g., $N=5$ in Fig.~\ref{fig2:model}B). Bumps are marginally stable to shifts and so translationally invariant (Fig.~\ref{fig2:model}C). Unstable solutions act as separatrices between the stable bumps. Section \ref{sec:determ_C4} provides details on the stability analysis.

Two limits of Eq.~(\ref{stair}) are of interest from previous studies of attractor solutions to Eq.~(\ref{eq:model_C4}). First, taking $N=1$, we recover a Heaviside nonlinearity, imposing a model with all-or-none responses, either exhibiting stable bumps or no activity, as shown by Amari~\cite{Amari77}. This limit has been useful in analyses of the dynamics of bumps as it allows for explicit calculation of solutions, localization of linear stability calculations, and interface methods for determining nonlinear dynamics~\cite{Amari77,coombes2004evans,Coombes2012InterfaceJNeuro,Krishnan18}. Alternatively, fixing $\theta_k = \theta \cdot k/N$ with $k\in \{1, 2, \dots, N\}$ and taking $N \to \infty$ generates a piecewise linear firing rate function
\begin{align*}
    f(u) = \begin{cases} 1, & u \geq \theta, \\ u/ \theta, & 0 < u < \theta, \\ 0, & u \leq 0.  \end{cases}
\end{align*}
Selecting the gain $1/ \theta$ fine tunes the model~\cite{lim2013balanced} so it exhibits bumps with a continuum of amplitudes~\cite{carroll2014encoding}. Other continuous forms of firing rate function $f(u)$ could be obtained in the limit $N \to \infty$ with careful choices of $\theta_k$.

Network connectivity $w(x)$ is assumed to be shaped as the difference $w(x-y)= w_E(x-y) - w_I(x-y)$, collapsing excitation and inhibition into a single population, which can be done rigorously using a separation of timescales analysis~\cite{Amari77,Pinto2001b}. Contributions from excitatory and inhibitory populations are given by von Mises distributions $w_k(x)=A_k  {\rm exp} \left[ \kappa_k \left[ \cos(x) - 1 \right] \right]$. Approximation of the effective weight function $w(x) = w_E(x) - w_I(x)$ using a finite set of even Fourier modes allows us to write
\begin{align*}
    w(x)=W_0+\sum_{j=1}^M W_j\cos(jx),
\end{align*}
where the first mode ($j=1$) is dominant and has positive weight, due to the local excitation and lateral inhibition~\cite{GoldmanRakic1995,Compte2000,Wimmer14}.

Cue contrast, size, and clarity is parameterized by the convolution:
    \begin{align}
        I_{c}(x,t)&=\frac{A_c(t)}{2} w(x)*\left[ {\rm erf} \frac{x+a_c}{\sigma_c} - {\rm erf} \frac{x-a_c} {\sigma_c} \right], \label{eq:cue}
\end{align}
where $A_c(t) = A_{\rm cue} \mathbb{I}_{[t^{c}_{\alpha}, t^{c}_{\omega}]}(t)$ describes the temporally dependent strength of the cue, $\mathbb{I}_{\Omega}(t) = 1$ for $t\in \Omega$ (cue on) and 0 otherwise (indicator function), and $a_{c}$ is the cue halfwidth. Increasing $\sigma_c \geq 0 $ smooths the input so that in limits $\sigma_c\rightarrow \infty$ flattens the profile and $\sigma_c\rightarrow 0^+$ yields a top hat convolved with the weight kernel.

Spatially-extended Wiener process increments have zero mean, spatial correlations $\langle dW(x,t)dW(y,z)\rangle = C(x-y)\delta(t-s)dtds$ with $\delta(x)$ the Dirac delta distribution, and are scaled to be weak ($0 < \epsilon \ll 1$). Spatial correlations are simulated by spatially convolving white noise increments $d\Upsilon(x,t)$ with an appropriate filter ${\mathcal F}(x)$, so that if $d W(x,t) = {\mathcal F}(x)*d \Upsilon(x,t)$, it can be shown that $C(x-y) = \int_{- \pi}^{\pi} {\mathcal F}(x-z) {\mathcal F}(y-z) dz$~\cite{Kilpatrick13WandBumpSIAD}. 

Numerical simulations (See Appendix~\ref{app:sims} for details) show cues of increasing salience (e.g., strength, time-length, size) generate bumps of increasing amplitude (Fig.~\ref{fig2:model}B). We next derive conditions for bumps, their stability, and their phase-amplitude dynamics in response to perturbations.

\section{\label{sec:determ_C4}Deterministic analysis}
Explicit bump solutions to Eq.~(\ref{eq:model_C4}) can be directly constructed using self-consistency. Stability is determined by an associated linearized operator. An appropriate ansatz inspired by observations from stability calculations then paved the way for low-dimensional reductions of bump dynamics to a set of evolution equations. We conclude this section by identifying how our metastable neural circuit models provide more robust representations of certainty and input angle than past models.

\subsection{Stationary solutions\label{ss:stability_C4}}
\begin{figure*}[t!]
\includegraphics[width=0.85\textwidth]{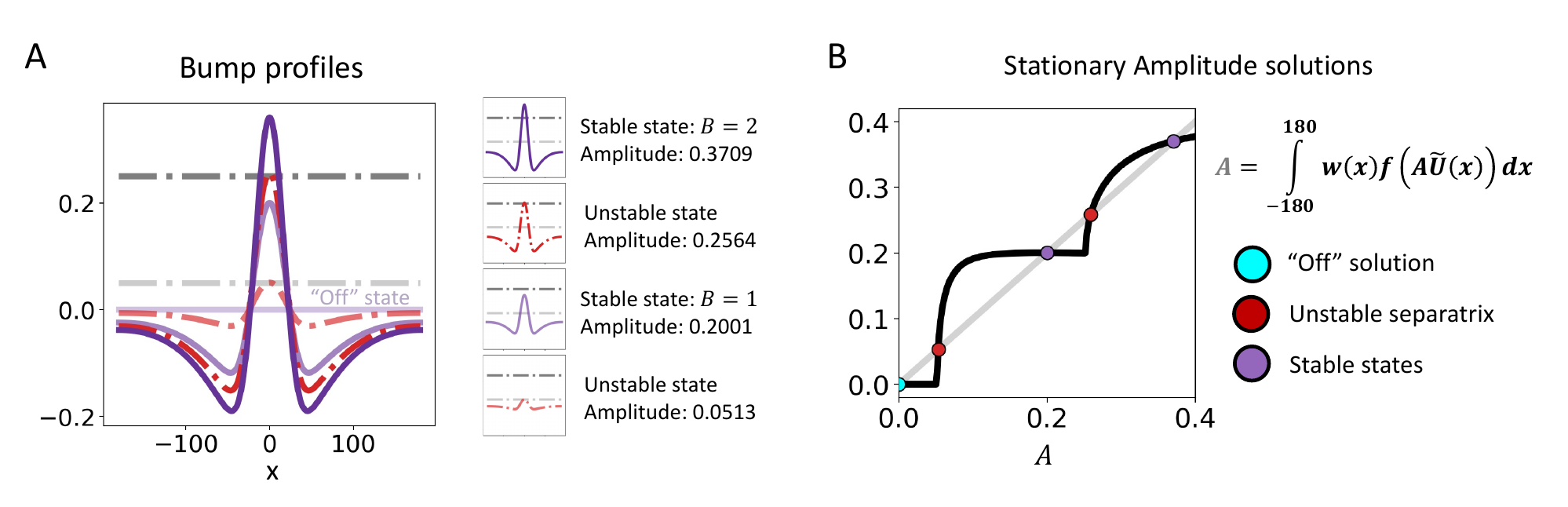}% Here is how to import EPS art
\caption{\textbf{Stationary bump solutions.} {\bf A.}~All five possible solutions to Eq.~(\ref{eq:model_C4}) given a staircase firing rate Eq.~(\ref{stair}) with $N=2$ are plotted $U(x)$ in degrees $x$: Stable ``off" state $U \equiv 0$; stable/unstable $B=1$ bump profiles (purple/red) only intersect lower threshold $U(a_1) = \theta_1$ (light dash dot line); stable/unstable $B=2$ bump profiles intersects low $U(a_1) = \theta_1$ and high $U(a_2) = \theta_2$ (dark dash dot line) thresholds. {\bf B.}~Bump solutions $U(x)$ all have roughly the same normalized profile ($\tilde{U}(x) = U(x)/U(0)$) allowing us to represent them by near-exact amplitude solutions to the implicit equation $A = \langle w(x), f(A \tilde{U}(x)) \rangle$, revealing the ``off" state (blue); unstable bumps (red); and stable bumps (purple) as points along the line of amplitudes $A$.}

\label{fig: twostep_stationary}
\end{figure*}

Time-independent solutions $u(x,t)=U(x)$ to Eq.~(\ref{eq:model_C4}) with $\epsilon \equiv 0$ and $I_{c} \equiv 0$ satisfy $U(x) = w(x)*f(U(x))$. Decomposing the weight function into $M$ Fourier modes, leveraging trigonometric identities, and examining even solutions, we find stationary solutions take the form:
\begin{align}
    U(x)&=\sum_{j=0}^M \underbrace{W_j \langle \cos(jx), f(U(x)) \rangle}_{\bar{U}_j} \cos(jx), \label{bumpcos}
\end{align}
where $\langle p(x), q(x) \rangle = \int_{- \pi}^{\pi} p(x) q(x) dx$ is an inner product. For any firing rate function, we can form a dense, nonlinear, implicit system for the coefficients $\bar{U}_j$~\cite{veltz2010local,Kilpatrick13WandBumpSIAD,bressloff2019stochastic}
\begin{align}
    \bar{U}_j = W_j \left\langle \cos (jx), f \left[ \sum_{j=0}^M \bar{U}_j \cos (jx) \right] \right\rangle. \label{Uj}
\end{align}
For the staircase firing rate $f$, Eq.~(\ref{stair}), we can find $N$ thresholds ($\theta_1 < \cdots < \theta_N$) such that there are $N$ possible bump solutions. Index bump states as $B = 1,...,N$ (e.g., $B=1$ and $B=N$ represent the lowest and highest bump amplitude states), then there are $B$ interfaces (or halfwidths) $a_i$ satisfying the level set conditions, $U(\pm a_i)=\theta_i$. The profile of the $B$th bump crosses $B$ levels of the firing rate function, where $1\leq B\leq N$, so stationary bumps satisfy
\begin{align*}
    U(x)=\frac{2}{N} \sum_{k=1}^B \left[ W_0 a_k+ \sum_{j=1}^M \frac{W_j\cos(jx)}{j} \sin(ja_k) \right].
\end{align*}
Utilizing the threshold-crossing conditions $\theta_i=U(a_i)$ for $i = 1,...,B$ one can implicitly define the half-widths $a_i$ from the system of equations
\begin{align}
\theta_i &=\frac{2}{N} \sum_{k=1}^B \left[W_0 a_k+ \sum_{j=1}^M \frac{W_j\cos(ja_i)}{j} \sin(ja_k) \right] \label{threshcond}
\end{align}
for $i = 1,...,B$. The system Eq.~(\ref{threshcond}) can be numerically solved iteratively across a range of thresholds (See Appendix~\ref{iterexist}). Cascades of saddle node bifurcations for each half-width and threshold pair emerge (See Fig.~\ref{fig:iterthresh} in Appendix~\ref{iterexist}).
Up to $N+1$ stable solutions (including the quiescent state, $U \equiv 0$) exist for a neural field with an $N$-step staircase firing rate, separated by $N$ unstable bumps (See Fig.~\ref{fig: twostep_stationary}A for $N=2$ example). Alternatively, one can also utilize approximations to stationary bump solutions assuming they are parameterized by a single amplitude $A$ which represents the scaling of the peak (Fig.~\ref{fig: twostep_stationary}B).

\subsection{Stability}
Stability of bumps can be determined by examining the linear dynamics of perturbations at the interfaces defined by the level sets $u(x,t) = \theta_i$ analogous to \cite{Pinto2001b,Amari77,Kilpatrick13WandBumpSIAD, cihak2024multiscale,coombes2004evans}. We study small smooth perturbations of the bump using the ansatz $u(x,t)=U(x)+\psi(x,t)$ 
where $||\psi||<<1$. 
For a given bump solution state $1\leq B\leq N$, we plug in the ansatz, Taylor expand, and truncate to first order to obtain the linearized dynamics
\begin{align}
    \partial_t\psi = -\psi + \frac{1}{N}\sum_{k=1}^B \sum_{a = \pm a_k} \frac{\psi(a)w(x-a)}{|U'(a_k)|} \equiv {\mc L} \psi \label{linstab}
\end{align}
where we define the linear operator
\begin{align*}
    {\mc L} u(x) = -u(x) + w(x)*\left[ f'(U(x)) u(x) \right].
\end{align*}
Note, to obtain this result, we have formally Taylor expanded the Heaviside nonlinearities that comprise $f(u)$ and whose discontinuities are shielded by integration against the perturbations $\psi (x,t)$. Formulas for the distributional derivatives are obtained by noting
\begin{align*}
    \delta(x+a_k) - \delta(x-a_k) &= \frac{d}{dx} \left[ H(x+a_k)-H(x-a_k) \right] \\
    &= \frac{d}{dx} H(U(x)-\theta_k) \\
    &= H'(U(x)-\theta_k) U'(x)
\end{align*}
and then dividing by the odd function $U'(x)$. Summing, we then find
\begin{align*}
    f'(U(x)) = \frac{1}{N} \sum_{k=1}^B \frac{\delta(x-a_k)+\delta(x+a_k)}{|U'(a_k)|},
\end{align*}
where we can determine
\begin{align*}
    U'(x) = \frac{1}{B} \sum_{k=1}^B \left[ w(x+a_k)-w(x-a_k) \right].
\end{align*}
Separating solutions $\psi(x,t)=\psi(x)e^{\lambda t}$ and evaluating Eq.~(\ref{linstab}) at interfaces $x = \pm a_1,\dots,\pm a_B$ localizes the stability problem to a discrete eigenvalue/vector system given by a $2B \times 2 B$ matrix.
The quiescent solution $u=0$ is stable (Fig.~\ref{fig2:model}D), due to the pure linear decay arising when $\psi(\pm a_k)\equiv 0$ for all $k$ in Eq.~(\ref{linstab}). For each $B$ where bump solutions exist, we generally find two stationary solutions: a stable wide solution and an unstable narrow solution that is a separatrix between the wide solution and the state below~\cite{Amari77,Kilpatrick13WandBumpSIAD}, finding no more than $N+1$ stable solutions and $N$ unstable solutions arising due to amplitude quantization of the metastable $N$ step firing rate function.

To illustrate how the stairstep firing rate function impacts the stability problem beyond the standard single step ($N=1$) case~\cite{Amari77}, consider $N=2$, so the linearized and localized eigenproblem becomes
\begin{align}
    (\lambda + 1) \psi(x) = \frac{1}{2} \sum_{k=1}^2 \sum_{a = \pm a_k} \frac{\psi(a) w(x-a)}{|U'(a_k)|}
\end{align}
for $x=\pm a_1, \pm a_2$, a $4 \times 4$ system. As expected, the bump is marginally stable to shifts. Assuming $\psi(-a_k) = - \psi(a_k)$ for $k=1,2$, plugging in $\lambda =0$, and enforcing self-consistency, we obtain a single equation relating perturbations of the inner $\psi(a_2)$ and outer $\psi(a_1)$ interfaces
\begin{align*}
    \frac{\psi(a_1)}{\psi(a_2)} = \frac{w(0) - w(2a_1) + w(\Delta a) - w(a_+)}{w(0) - w(2a_2) + w(\Delta a) - w(a_+)},
\end{align*}
where $\Delta a = a_2 - a_1$ and $a_+ = a_1+a_2$. This reflects the marginal stability due to translation invariance. We do not expect general bump stability conditions to emerge from examining these perturbations. Stability is often determined by studying width perturbations which are even symmetric $\psi(-a_k) = \psi(a_k)$  for $k=1,2$. This generates a $2 \times 2$ eigenproblem whose solutions imply stability given positive determinant and negative trace of the associated matrix, providing the conditions
\begin{align*}
    & \hspace{3mm} 2 w(2a_1) w(2a_2) > (w(2a_1)+w(2a_2))(w(\Delta a)-w(a_+)) \\
    & \hspace{28mm} {\rm and} \\
    &(w(\Delta a)-w(a_+))^2 > - 2 w(2a_1) w(2a_2) + \\
    & \hspace{15mm} 2(w(2a_1)+w(2a_2))(w(0)+w(\Delta a) - w(a_+)).
\end{align*}
Alternative conditions can be constructed for other perturbation types, including cases where interfaces at different levels are shifted in opposite directions.

\subsection{\label{ssec: ampl_approx}Reduced equation for amplitude evolution}
\begin{figure}[t!]
    \centering
    \includegraphics[width=0.5\textwidth]{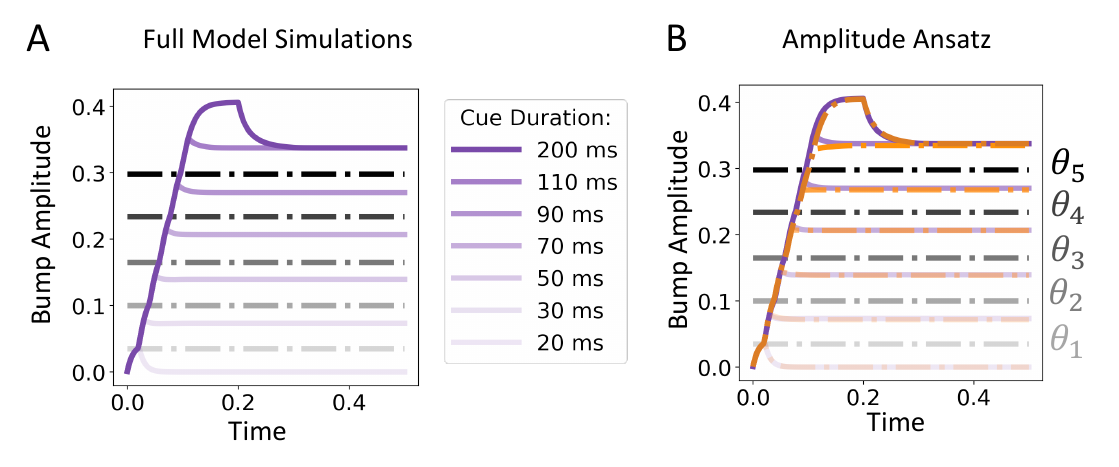}
    \caption{\textbf{Stimulus-driven bump amplitude dynamics.} {\bf A.}~Increasing cue duration in Eq.~(\ref{eq:model_C4}) generates larger activity responses, which settle into larger amplitude bumps. {\bf B.}~
    Amplitude ansatz Eq.~(\ref{phaseampansatz}) and the amplitude evolution Eq.~(\ref{loneamp}) approximate the build-up and relaxation of the bump amplitude in the full Eq.~(\ref{eq:model_C4}) well. Mild inaccuracies arise due to the assumption of fixed shape $u(x,t)/u(0,t) \approx \tilde{U}(x)$.}
    \label{fig:coeff_ampl_evolution}
\end{figure}

Low-dimensional reductions of neural field dynamics on the ring $x \in [- \pi, \pi)$ can be derived using Fourier decompositions~\cite{ben1995theory,ermentrout1998neural,Laing2001,veltz2010local,Kilpatrick13WandBumpSIAD,bressloff2019stochastic}. A complementary approach uses eigenfunctions of the linearized system to partition dynamics into a position variable for a bump (phase) and its amplitude~\cite{carroll2014encoding,kilpatrick2016ghosts}. Such an approach starts with the ansatz
\begin{align}
    u(x,t) = A(t) \tilde{U}(x-\Delta(t)) + \psi (x- \Delta(t),t), \label{phaseampansatz}
\end{align}
with $||\psi|| \ll 1$, assuming to leading order that perturbations of the bump shift its amplitude $A(t)$ and phase $\Delta(t)$ but otherwise the bump roughly retains its shape defined as $\tilde{U}(x) = U(x)/U(0)$. A low-dimensional description of input-driven bump dynamics can then be obtained by first plugging Eq.~(\ref{phaseampansatz}) into a noise-free version of Eq.~(\ref{eq:model_C4}), expanding and truncating to obtain leading order terms:
\begin{align*}
          A'  \tilde{U}(x) - \Delta' A \tilde{U}'(x) \approx& -A \tilde{U}(x) + w(x)*f(A\tilde{U}(x)) \\
          & + I_c(x+ \Delta, t).
\end{align*}
To obtain evolution equations for the phase and amplitude $(\Delta, A)$, we exploit even (odd) symmetry of $\tilde{U}(x)$ ($\tilde{U}'(x)$) to isolate the temporal derivative $A'(t)$ ($\Delta'(t)$) by taking inner products and rearranging
\begin{subequations} \label{phaseamp}
\begin{align}
    A' &= - A + G(A) + J_A(\Delta, t), \label{phaseampA}\\
    \Delta' &= -\frac{1}{A}  J_{\Delta}(\Delta, t), \label{phaseampdel}
\end{align}
\end{subequations}
where $||p||^2 = \langle p(x), p(x) \rangle$ is the squared norm induced by the innner product and
\begin{align*}
    G(A) = \frac{\langle \tilde{U}(x), w(x)*f(A \tilde{U}) \rangle}{||\tilde{U}||^2}
\end{align*}
describes the impact of recurrent connectivity on the bump amplitude, and
\begin{align*}
    J_A(\Delta, t) &= \frac{\langle \tilde{U}(x), I_c(x + \Delta, t) \rangle}{||\tilde{U}(x)||^2}, \\
    J_{\Delta} (\Delta, t) &= \frac{\langle \tilde{U}'(x), I_c(x+\Delta, t) \rangle}{||\tilde{U}'(x) ||^2},
\end{align*}
describe how the even and odd parts of the cue input steer the amplitude and phase. The phase is shifted by cues that apply odd perturbations to the bump, though increasing the amplitude $A$ of the bump decreases these shifting responses. 
Amplitudes relax to a stable steady state once cues are shut off, determined by the basin of attraction demarcated by $\theta_i$ where they reside (Fig.~\ref{fig:coeff_ampl_evolution}A). Changing cue contrast, size, and clarity also alters long term bump amplitudes (See Appendix \ref{app:cue_char} and Fig.~\ref{fig:features_vs_amplitudes}). Assuming separability of the cue $I_c(x,t) = I_A(t) {\mc J}(x)$, the phase $\Delta$ in Eq.~(\ref{phaseamp}) will not shift, so taking $\Delta(0) = 0$ without loss of generality, we can reduce the system to
\begin{align}
    A' &= - A+ G(A) + \bar{\mc J}  I_A(t), \label{loneamp}
\end{align}
where $\bar{\mc J} = \langle \tilde{U}(x), {\mc J}(x) \rangle/ ||\tilde{U}(x)||^2$. Dynamics of Eq.~(\ref{loneamp}) match the build up and relaxation to steady state amplitudes determined from full simulations with low error (Fig.~\ref{fig:coeff_ampl_evolution}B). We can thus use Eq.~(\ref{loneamp}) to approximate the transient dynamics and stable bump profiles expected (freezing $I_A(t) \equiv \bar{I}_A$) as the input amplitude is varied.

\begin{figure*}
    \centering
    \includegraphics[width=0.84\textwidth]{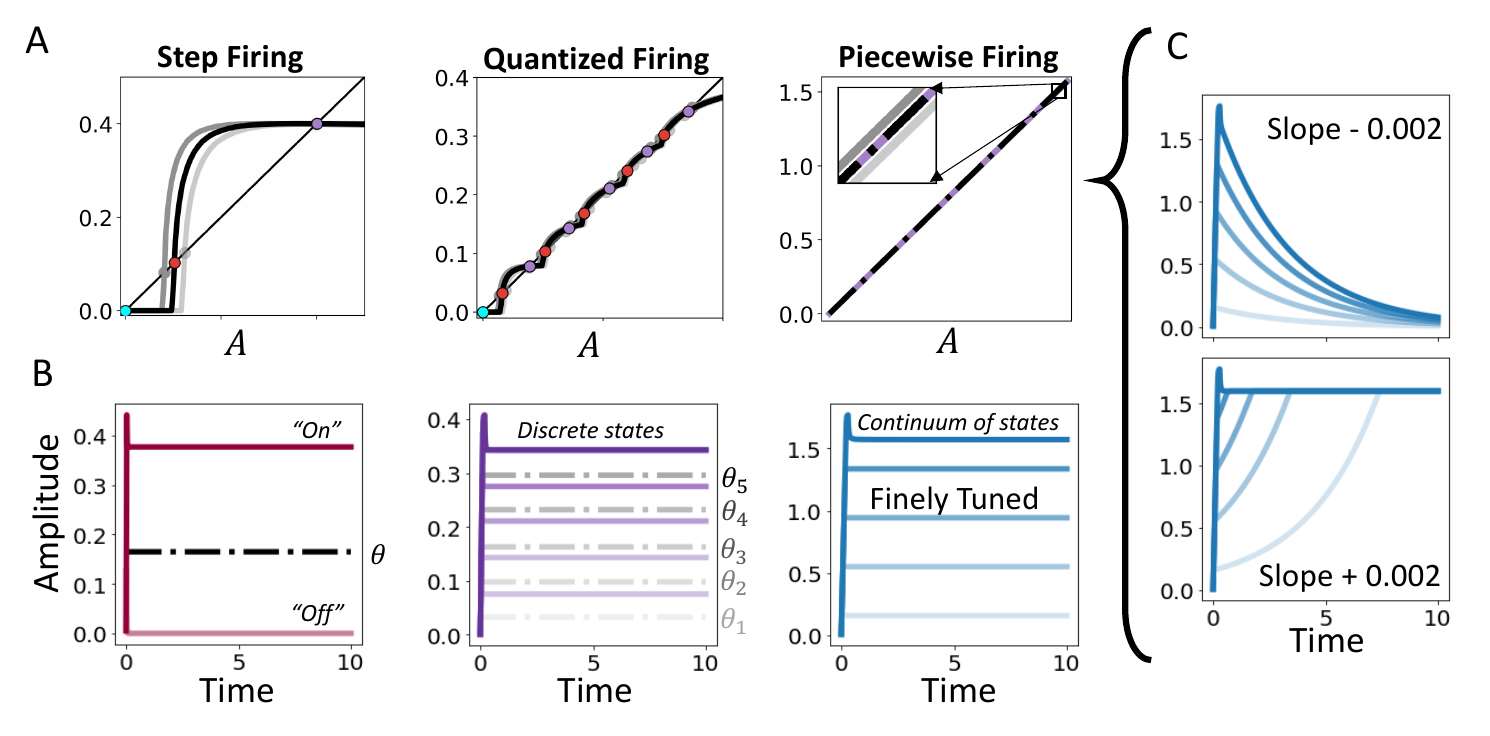}

    \caption{\textbf{Metastable neural circuit robustly encodes amplitude.} {\bf A.}~Memory robustness and flexibility tradeoff in circuits with quantized firing rates. Amplitudes of stationary bumps $A = \max_x U(x)$ are represented as intersections (dots) of the inner product $\langle w, f(A \tilde{U}) \rangle$ (thick line) and line of unity (thin line). Single step nonlinearities $f(u) = H(u-\theta)$ support one stable bump, while staircase nonlinearities, Eq.~(\ref{stair}), with $N$ steps can support $N$ stable bumps. Equilibria persist in both models even when $f(u)$ is perturbed (grey lines). A finely tuned piecewise linear firing rate supports a continuum of bumps, but perturbations (see inset for slope perturbations) annihilate the line attractor. {\bf B.}~Inputs of varying time lengths either lead to a single active bump amplitude or no activity for the Heaviside network, a discrete and graded set of bumps amplitudes for the staircase network, and a set of graded amplitudes along a continuum for the piecewise linear network~\cite{carroll2014encoding}. The Heaviside and staircase networks maintain these solutions under model perturbations.
    {\bf C.}~Perturbations to the piecewise linear firing rate function lead to bump collapse or the trivial quiescent solution, breaking the amplitude coding of the finely tuned system.}
    \label{fig:C4_robust}
\end{figure*}

\subsection{Bump robustness to model perturbations}
Activity states in metastable neural circuits are robust to dynamic perturbations, and also structural perturbations like changes to connectivity or firing rate relations~\cite{goldman2003robust}. Line attractor models which finely encode stimulus differences can be generated by considering piecewise linear firing rate relations~\cite{machens2005flexible}, as in hand-designed neural circuit models with a continuum of bump attractor amplitudes~\cite{carroll2014encoding,kutschireiter2023bayesian,esnaola2022flexible}. Their low-dimensional dynamics lie upon a planar attractor whose angular direction encodes stimulus estimates and radial dimension represents estimate certainty. However, even mild structural perturbations (Fig~\ref{fig:C4_robust}A) destroy the carefully crafted continuum of amplitudes (Fig.~\ref{fig:C4_robust}C), motivating more robust representations. Commonly used single step (Heaviside) nonlinearities in $f(u)$ in Eq.~(\ref{eq:model_C4}) can only support bumps with a single amplitude or a quiescent state, depending on the duration of cues (Fig.~\ref{fig:C4_robust}B), but bump solutions are more robust. 

Our intermediate solution balances robustness and flexibility by considering staircase firing rate functions, Eq.~(\ref{stair}), retaining multiple stable bump amplitude states even when structurally perturbed (Fig.~\ref{fig:C4_robust}A,B). Even weak cues can generate bumps, which do not arise in the single step case. On the other hand while the piecewise linear firing rate supports a continuum of possible amplitude states for different cue durations (Fig.~\ref{fig:C4_robust}B), infinitesimal structural perturbations (e.g., slope/threshold changes to the firing rate function, connectivity perturbations) annihilate the line attractor  (Fig.~\ref{fig:C4_robust}C). Whereas the quantized firing rate function allows for robustness to structural perturbations while still providing an appreciable resolution of stimulus representations.

% It is important to note that similar to how the step-like feedback bands needing to balance the decay in a network with dendritic hysterisis in ~\cite{goldman2003robust} the staircase does need to be tuned to achieve $N$ amplitude states. However, the choice of parameters is still more robust to slight mistuning, far moreso than the linear firing rate where even the slightest perturbation result in a collapse of the system. 

\section{\label{sec:stochas_C4}Stochastic dynamics of bump phase and amplitude}
\begin{figure*}[t!]
    \centering
    \includegraphics[width=0.85\textwidth]{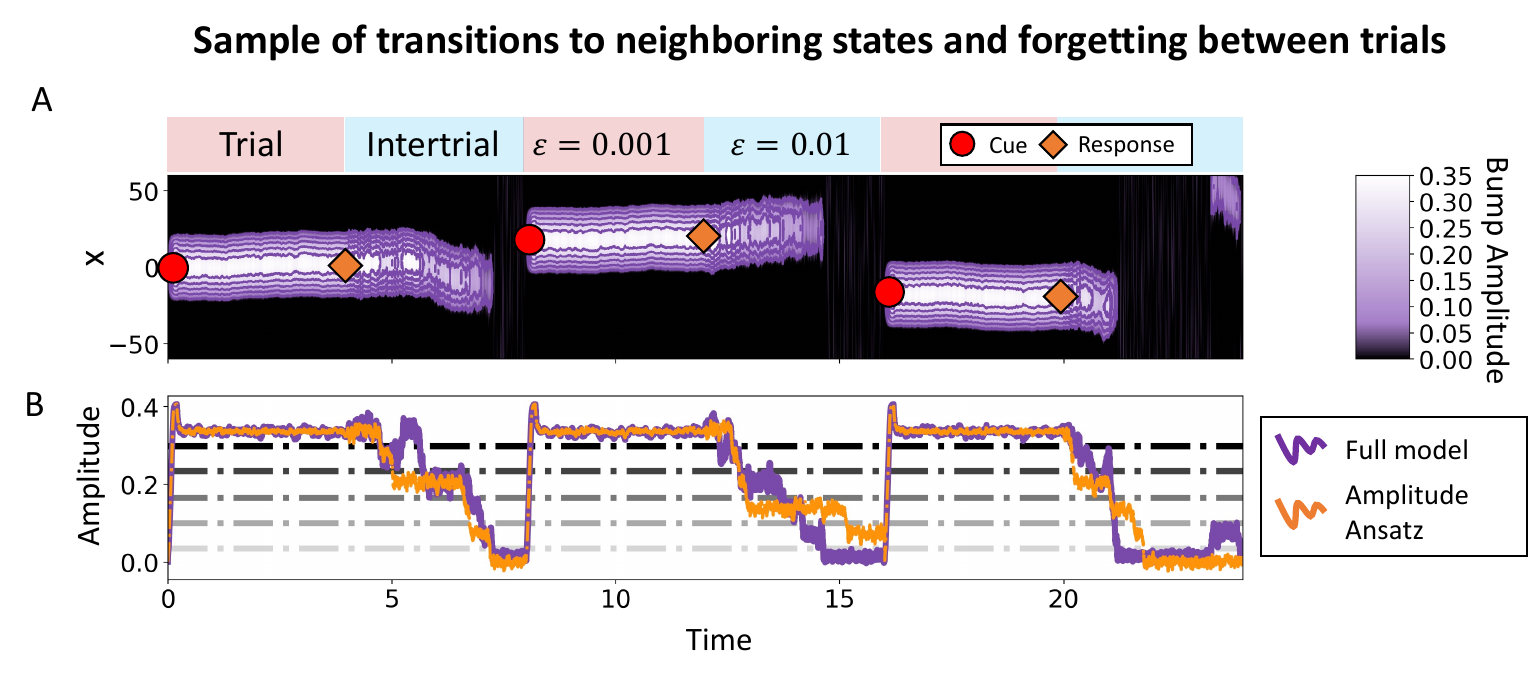}
    \caption{\textbf{Bump amplitude transitions.} {\bf A.}~Noise ($\epsilon=0.001$) perturbs neural activity (purple heatmap) so the bump wanders within a trial. Note, level sets (purple lines) $u=\theta_i$. The reduction of attention between trials is modeled by an increase in noise ($\epsilon=0.01$) which rapidly drives the bump to extinction through transitions in amplitude states. {\bf B.}~Comparison of amplitude dynamics identified in the full model simulation (purple) Eq.~(\ref{eq:model_C4}) and amplitude ansatz (orange) Eq.~(\ref{lonestochamp}).}
    \label{fig:transition_samples}
\end{figure*}

Responses from tasks requiring delayed estimates of continuum quantities have been reliably modeled by bump attractor models and their low-dimensional approximations~\cite{kilpatrick2013optimizing,Wimmer14,Constantinidis16,panichello2019error}. The phase $\Delta (t)$ (e.g., centroid or peak) of the bump encodes the estimate~\cite{Wimmer14}, so the phase variance $\langle (\Delta (t)-\Delta(0))^2 \rangle $ across trials models memory degradation~\cite{burak2012fundamental,Kilpatrick13WandBumpSIAD} and scales linearly with delay time~\cite{White94,Ploner98} (See however \cite{bliss2017serial,panichello2019error} for more complex accounts of memory degradation). Strengthening cues in our model increases the salience of bumps and the estimates they encode. Noise in Eq.~(\ref{eq:model_C4}) causes bumps to wander diffusively with larger amplitude bumps wandering less~\cite{Kilpatrick13WandBumpSIAD,carroll2014encoding,Krishnan18}, and bump amplitudes can transition to neighboring values (Fig.~\ref{fig:transition_samples}). We can derive accurate estimates of the rate of these transitions, providing a new and extended theory of the degradation of delayed estimate accuracy in neural circuits. Our reduced phase-amplitude equations can also be used to estimate phase variance across all possible bump amplitudes. 

\subsection{Stochastic phase-amplitude equations}

In the analogous deterministic system, we showed the ansatz Eq.~(\ref{phaseampansatz}) decomposes the effects of odd (even) perturbations into shifts (scalings) of the bump. Stochastic perturbations from the spatially-extended Wiener process noise in Eq.~(\ref{eq:model_C4}) generate wandering in the phase variable $\Delta (t)$~\cite{Kilpatrick13WandBumpSIAD}, and occasional transitions in bump amplitude $A(t)$ to neighboring attractors. Plugging in the ansatz Eq.~(\ref{phaseampansatz}) and integrating against the even and odd functions $\tilde{U}(x)$ and $\tilde{U}'(x)$, we find a coupled system of stochastic differential equations
\begin{subequations} \label{stochphaseamp}
\begin{align}
    d A  =& \left[ -A + G(A) + J_A(\Delta, t) \right] dt + \sqrt{\epsilon}dZ_{A}(\Delta, t), \label{sphaseampa} \\
    d \Delta =& -\frac{1}{A} J_{\Delta}(\Delta, t)dt - \frac{\sqrt{\epsilon}}{A} d Z_{\Delta}(\Delta, t), \label{sphaseampb}
\end{align}
\end{subequations}
where noise increments are obtained by separating even and odd parts
\begin{align*}
    dZ_{A}(\Delta,t) &= \frac{\langle \tilde{U}(x), d W(x+\Delta, t) \rangle}{||\tilde{U}(x)||^2}, \\
    dZ_{\Delta}(\Delta,t) &= \frac{\langle \tilde{U}'(x), d W(x+\Delta, t) \rangle}{||\tilde{U}'(x)||^2}.
\end{align*}
Increasing the bump amplitude dampens the impact of perturbations on the phase. Eq.~(\ref{stochphaseamp}) describes the stochastic dynamics of the bump phase and amplitude, accounting for non-equilibrium dynamics of the amplitude $A$ (See also \cite{kilpatrick2016ghosts}). Amplitude dynamics in Eq.~(\ref{sphaseampa}) can be further approximated by a quantized chain of Markovian states assuming amplitudes remain near equilibria until fluctuations kick them to neighboring steady states of the deterministic system (roots of $A = G(A)$)~\cite{lindner2001optimal,Kilpatrick13WandBumpSIAD}. The phase $\Delta$ lies on a continuum ring attractor $[- \pi, \pi)$, wandering with a diffusion coefficient determined by the bump amplitude $A$. We leverage our phase-amplitude system to estimate the mean time to transition between amplitude states (Fig.~\ref{fig:transition_samples}), which impacts the wandering of bump phase and the estimate retention.

\subsection{Mean time for amplitude transitions}
\begin{figure}[t!]
    \centering
    \includegraphics[width=0.475\textwidth]{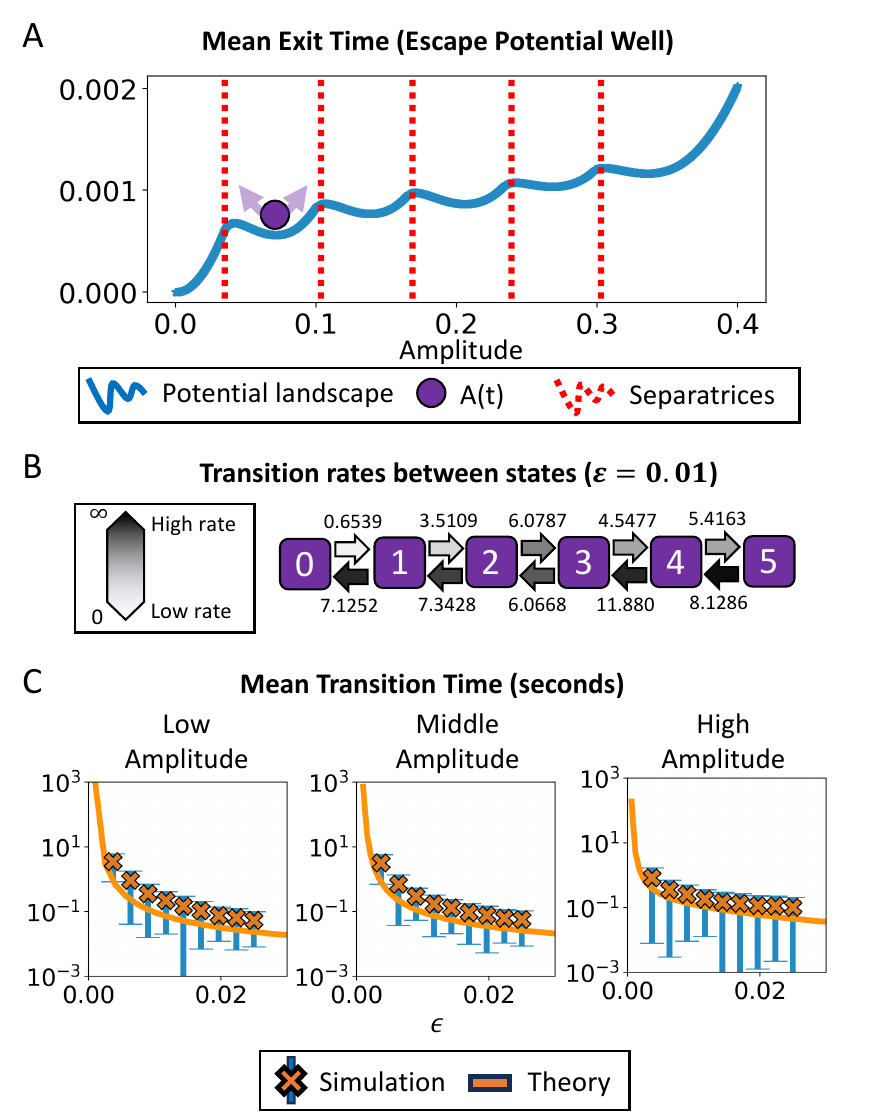}
    \caption{{\bf Amplitude potential landscape and transition dynamics.} {\bf A.}~Amplitude potential well landscape $V(A)$ (blue line) determines drift via the descent of its gradient $-V'(A)=-A+G(A)$. Potential peaks (red dashed) separate stable minima. Stochastic fluctuations drive amplitude (purple particle) to escape minima, usually {\em downhill} towards the off state ($A \equiv 0$). {\bf B.}~Markov chain approximation of well hopping dynamics. Transition rates from a state $\bar{A}_i^s$ to its neighbor $\bar{A}_{i \pm 1}^s$ are approximated $r_i^{\pm} \approx \pi_i^{\pm}(\bar{A}_i^s)/{\mc T}(\bar{A}_i^s)$ by the ratio of the escape probability and mean first passage time. Stable bump amplitudes are enumerated from $0$ (the off state) and $5$ (the highest amplitude state). Noise $\epsilon=0.01$.  \textbf{C.}~Mean transition times were estimated by averaging over 1000 simulations (Mean: orange `x'; blue lines: standard deviation) comparing well with theory (orange line). See Appendix \ref{iterexist} for simulation details.}
\label{fig:potential_transitions}
\end{figure}

Defining stable ($\bar{A}_i^s$) and unstable ($\bar{A}_i^u$) bump amplitudes of the noise-free system ($A = G(A)$), we have observed (Fig.~\ref{fig:transition_samples}) that the full system tends to dwell near stable amplitudes ($\bar{A}_i^s$) on short timescales, eventually hopping to neighboring values ($\bar{A}_{i\pm1}^s$). In Eq.~(\ref{stochphaseamp}), the amplitude $A$ must pass through unstable bump amplitudes ($\bar{A}_i^u$ or $\bar{A}_{i+1}^u$) when transitioning. Between transitions and in the absence of inputs, the translation symmetry of the spatially-extended Wiener process statistics ensures Eq.~(\ref{sphaseampa}) behaves as a one-dimensional stochastic differential equation
\begin{align}
    dA = \left[ - A + G(A) \right] dt + \sqrt{\epsilon} d \bar{Z}_A(t). \label{lonestochamp}
\end{align}
%The phase Eq.~(\ref{sphaseampb}) can be approximated by assuming the amplitude $A$ terms therein are piecewise constant in time, taking only the values $\bar{A}_i^s$, determining the impacts of noise when adjacent to the $i$th stable bump amplitude.
Fluctuation-induced transitions in amplitude $A$ are determined by analyzing the associated Fokker-Planck equation of Eq.~(\ref{lonestochamp}). We can then formulate the mean exit time problem for $A(t)$ to depart the interval $[\bar{A}_i^u,\bar{A}_{i+1}^u]$ when starting at $\bar{A}_i^s$ ($i=0,1,...,N$). On the boundaries, $\bar{A}_0^u \to - \infty$ and $\bar{A}_{N+1}^u \to \infty$. The variance and diffusion coefficient of the noise in Eq.~(\ref{lonestochamp}) can be determined as $\langle \bar{Z}_A(t)^2 \rangle = D_A t$, where
\begin{align*}
    D_A = \frac{\int_{- \pi}^{\pi} \int_{- \pi}^{\pi} \tilde{U}(x) \tilde{U}(y) C(x-y) dy dx}{||\tilde{U}(x)||^4}.
\end{align*}
The probability density $p(A,t)$ evolves according to the Fokker-Planck equation
\begin{align}
    p_t &= - \frac{\pd}{\pd A} \left[ (-A+G(A)) p \right] + \frac{D_A}{2} p_{AA}, \label{fpeqn}
    %\\
    %&\equiv - {\mc J}_{A}(A,t), \nonumber
\end{align}
and $p(A,0) = \delta (A - \bar{A}_0)$, the amplitude starts at some value $A_0 \in [\bar{A}_i^u, \bar{A}_{i+1}^u]$. We expect $A_0 = \bar{A}_i^s$, but to determine first passage time statistics, we determine quantities across the interval. Since it determines the timescale on which a stationary approximation of $A$ in Eq.~(\ref{sphaseampb}) is valid, as well as the higher order dynamics of $A$, we are interested in the random time $T(A_0)$ the amplitude in Eq.~(\ref{lonestochamp}) escapes the interval $[\bar{A}_i^u, \bar{A}_{i+1}^u]$. The mean time ${\mc T}(A_0) = \langle T(A_0) \rangle$ is determined by leveraging the backward Fokker-Planck (FP) equation~\cite{Gardiner2009}, describing the evolution of the probability $q \equiv p(A,t|A_0,0)$ we find the amplitude at $A$ at time $t$ given it started at $A_0$ at $t=0$. The state variable in the backward FP equation is the initial condition $A_0$ and we use the adjoint linear operator of Eq.~(\ref{fpeqn}) to define the flux
\begin{align}
    q_t &= \left[ -A_0+G(A_0) \right] q_{A_0} + \frac{D_A}{2} q_{A_0A_0}, \label{backfpeqn} \\
    &= -{\mc J}(A,t|A_0, 0). \nonumber
\end{align}
The probability we find the amplitude within $[\bar{A}_i^u, \bar{A}_{i+1}^u]$ at time $t$ is given by integrating the density
\begin{align*}
    \int_{\bar{A}_i^u}^{\bar{A}_{i+1}^u} p(A,t | A_0, 0) dA = {\mc G}(A_0, t) = P(T(A_0)>t),
\end{align*}
where the last equality follows from the fact that the amplitude leaves the interval after $t$ if it has not left by then. Integrating the backward FP Eq.~(\ref{backfpeqn}), we obtain a related equation for ${\mc G}(A_0,t)$
\begin{align}
    {\mc G}_t = \left[ - A_0 + G(A_0) \right] {\mc G}_{A_0} + \frac{D_A}{2} {\mc G}_{A_0 A_0}, \label{Geqn}
\end{align}
with boundary conditions ${\mc G}(A_0, 0) = 1$ if $A_0 \in [\bar{A}_i^u, \bar{A}_{i+1}^u]$ and 0 otherwise, while ${\mc G}(\bar{A}_i^u,t) = {\mc G}(\bar{A}_{i+1}^u,t) = 0$.  The mean first passage time can then be computed
\begin{align*}
    {\mc T}(A_0) = - \int_0^{\infty} t {\mc G}_t(A_0,t) dt = \int_0^{\infty} {\mc G}(A_0, t) dt.
\end{align*}
A differential equation for ${\mc T}(A_0)$ can be derived by integrating Eq.~(\ref{Geqn}) over $t \in (0, \infty)$, finding
\begin{align}
    \left[ - A_0 + G(A_0) \right] {\mc T}' + \frac{D_A}{2} {\mc T}'' = -1, \label{mfpteqn}
\end{align}
along with boundary conditions ${\mc T}(\bar{A}_i^u) = {\mc T}(\bar{A}_{i+1}^u) = 0$. Eq.~(\ref{mfpteqn}) can be solved by integrating to find
\begin{align*}
    {\mc T}(A) =& \frac{2}{D_A \Lambda(\bar{A}_i^u, \bar{A}_{i+1}^u)} \left[ \Lambda (\bar{A}_i^u, A_0)\int_{A_0}^{\bar{A}_{i+1}^u} \frac{{\mc V}(\bar{A}_i^u, y')}{ \nu (y')}dy' \right. \\
    & \hspace{1cm} \left. - \Lambda (A_0, \bar{A}_{i+1}^u) \int_{\bar{A}_i^u}^{A_0} \frac{{\mc V}(\bar{A}_i^u, y')}{\nu (y')} \right]
\end{align*}
where ${\mc V}(a,b) = \int_a^b \nu(x) dx$ and $\Lambda(a,b) = \int_a^b \frac{dx}{\nu(x)}$
and
\begin{align}
    \nu(A_0) = \exp \left[ \frac{2}{D_A} \left( V(\bar{A}_i^u)-V(A_0) \right) \right]. \label{exppot}
\end{align}
Stochastic amplitude dynamics are strongly determined by the potential function (Fig.~\ref{fig:potential_transitions}A), formed by integrating $V(A) = \int_{- \infty}^A \left[ A'- G(A') \right] dA'$, which biases transitions to lower amplitude states over time. The energy barrier the stochastic particle must surmount is lower on the left side (Fig.~\ref{fig:potential_transitions}A), so increases in neural variability following task-relevant epochs (when variability is lower, due perhaps to attention~\cite{mitchell2007differential,cohen2009attention}) could serve to annihilate persistent activity (See Fig.~\ref{fig:transition_samples} and \cite{Kilpatrick13WandBumpSIAD,kilpatrick2016ghosts}).

\begin{figure*}[t!]
    \centering
    \includegraphics[width=0.85\textwidth]{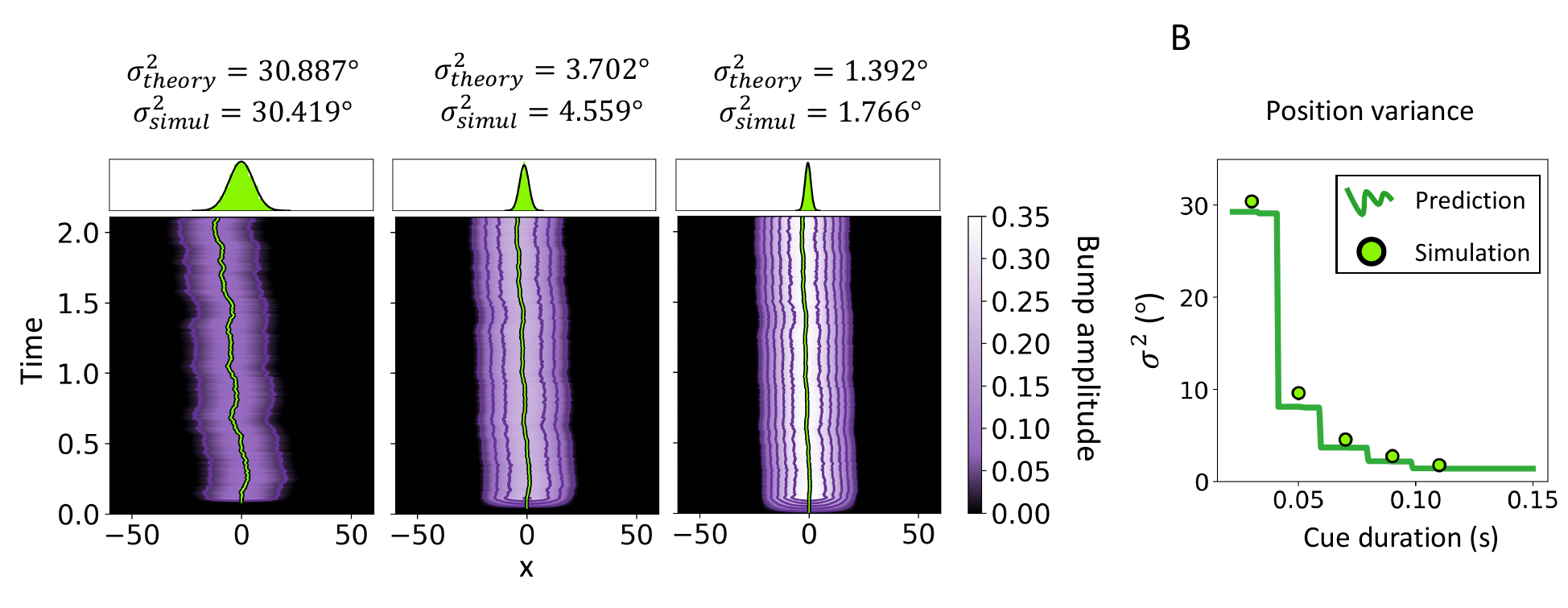}
    \caption[Increased cue duration decreases wandering.]{\textbf{Increased cue duration decreases wandering.}  {\bf A.}~Top: Estimate distributions computed from bump peak after delay for cue durations $t_c = 30, 70, 110$ms. Bottom: Wandering bumps $u(x,t)$ of low, medium, high amplitude following cues of different lengths. {\bf B.}~Response variance decreases with cue duration due to increasing bump amplitude. Noise $\epsilon = 0.01$. See Table~\ref{tableparam_C4} for other parameters and Appendix~\ref{app:sims} for numerical methods.} 
    \label{fig:wander_c4}
\end{figure*}

To approximate the rate of transition over either barrier, we assume that the mean time of escape over either boundary will be roughly the same ${\mc T}^{\pm}(\bar{A}_i^s) \approx {\mc T}(\bar{A}_i^s)$ as is often the case even with asymmetric potentials~\cite{lindner2001optimal,Gardiner2009}. We then approximate the rate of transition over either barrier $r_i^{\pm} \approx \pi_i^{\pm}(\bar{A}_i^s)/{\mc T}(\bar{A}_i^s)$ as the escape probability scaled by the mean time. Exit probabilities $\pi_i^{\pm}(\bar{A}_i^s)$ are determined by deriving the appropriate differential equation. First, integrate the probability current through the boundary of interest ${\mc J}(\bar{A}_{i+1}^u,t|A_0,0)$ or $-{\mc J}(\bar{A}_{i}^u,t|A_0, 0)$. For instance, the probability the particle exits via $A = \bar{A}_{i+1}^u$ after time with $t$ is
\begin{align*}
    g^+(A_0,t) &\equiv \int_t^{\infty} {\mc J}(\bar{A}_{i+1}^u, t' | A_0, 0) dt' \\
    &= \int_t^{\infty} \left[ \left( A_0 - G(A_0) \right) q - \frac{D_A}{2} q_{A_0} \right] dt'.
\end{align*}
Using the fact that $q = p(\bar{A}_{i+1}^u, t | A_0, 0)$ satisfies Eq.~(\ref{backfpeqn}), we find that $g_+(A_0,t)$ satisfies
\begin{align*}
g^+_t &= \left( -A_0 + G(A_0) \right) g^+_{A_0}+ \frac{D_A}{2} \pd_{A_0}^2 g^+_{A_0A_0}.
    %\int_t^{\infty} \pd_{t'} {\mc J}(\bar{A}_{i+1}^u, t' | A_0, 0) dt' \\
%    =& - {\mc J}(\bar{A}_{i+1}^u, t | A_0, 0) \\
%    =& \pd_t g_+ (A_0,t).
\end{align*}
Taking $t \to 0^+$ and defining $\pi_i^+(A_0) : = g^+(A_0,0)$, we see that ${\mc J}(\bar{A}_{i+1}^u, 0 | A_0, 0)$ vanishes if $A_0 \neq \bar{A}_{i+1}^u$, since $p(\bar{A}_{i+1}^u, 0 | A_0, 0) = \delta (A_0 - \bar{A}_{i+1}^u)$, so $g_t^+(A_0,0) \to 0$ and
\begin{align*}
   \left( -A_0 + G(A_0) \right) \pd_{A_0} \pi_i^{+}(A_0) + \frac{D_A}{2} \pd_{A_0}^2 \pi_i^+(A_0) = 0,
\end{align*}
where $\pi_i^+(\bar{A}_{i+1}^u) = 1$, $\pi_i^+(\bar{A}_i^u) = 0$, and $\pi_i^+(A_0) + \pi_i^-(A_0) = 1$. We solve and $\pi_i^+ (A_0) = {\mc N}(A_0)/ {\mc N} (\bar{A}_{i+1}^u)$ and $\pi_i^- (A_0) = 1 - \pi_i^+ (A_0)$ where ${\mc N}(A) = \int_{\bar{A}_i^u}^{A} \frac{dy}{\nu(y)}$.
%\begin{align*}
%    \pi_i^+(A_0) = \frac{\int_{\bar{A}_i^u}^{A_0} \frac{dy}{\nu(y)}}{\int_{\bar{A}_i^u}^{\bar{A}_{i+1}^u} \frac{dy}{\nu(y)}}, \ \ \ \pi_i^-(A_0) = \frac{\int_{A_0}^{\bar{A}_{i+1}^u} \frac{dy}{\nu(y)}}{\int_{\bar{A}_i^u}^{\bar{A}_{i+1}^u} \frac{dy}{\nu(y)}%\end{align*}
The escape probability and exit rate associated with the left boundary of each well is larger than for the right boundary (Fig.~\ref{fig:potential_transitions}B), so $A(t)$ will tend towards 0, and all bumps are eventually extinguished given a long delay time, as suggested by behavior~\cite{shin2017effects}.

Mean transition time estimates align well with full system simulations (Fig.~\ref{fig:potential_transitions}C), making two key predictions. First, higher amplitude bumps transition to neighboring amplitudes (usually lower) quicker than low amplitude bumps. This does not negate our overall claim that high amplitude bumps are more robust, since transitions from high amplitudes still generate medium to high amplitude bumps. Second, transition frequency increases with the fluctuation strength. Asymptotic errors in approximating the potential exponentially impact passage time estimates, which is obvious at low noise levels.

\subsection{Phase variance estimates}

We now study how bump amplitude shapes the wandering of the bump phase $\Delta (t)$. As has been found previously, higher amplitude bumps wander less~\cite{Kilpatrick13WandBumpSIAD,carroll2014encoding}. Since the phase encodes the remembered stimulus value $\Delta_0$, the variance $\langle (\Delta(t) - \Delta_0)^2 \rangle$ measures recall error. Variance is determined by analyzing the reduced and forced equation for the phase, Eq.~(\ref{sphaseampb}), which we write out in terms of integrals without external inputs
\begin{align}
    d \Delta = -\frac{\sqrt{\epsilon}}{A} \frac{\int_{- \pi}^{\pi} \tilde{U}'(x) d W(x + \Delta,t) d x}{\int_{- \pi}^{\pi} \tilde{U}'(x)^2 dx}. \label{noisephase}
\end{align}
Fixing the amplitude $A \approx \bar{A}_i^s$ in Eq.~(\ref{noisephase}) in the case of rare transitions due to weak noise and/or short delays, we compute variance
\begin{align*}
    \langle (\Delta - \Delta_0)^2 \rangle = \frac{\epsilon}{\bar{A}^2} \frac{\int_{- \pi}^{\pi} \tilde{U}'(x) \tilde{U}'(y) C(x-y) dy dx}{\left[ \int_{- \pi}^{\pi}\tilde{U}'(x)^2 dx \right]^2} t.
\end{align*}
Larger amplitudes $\bar{A}$ reduce the variance (Fig.~\ref{fig:wander_c4}), well predicted by our theory, as in findings showing higher certainty reduces response errors~\cite{honig2020humans,basso1997modulation}. Increased neural responses (bumps of higher amplitude) occur in response to longer, brighter, clearer, and larger cues generating more accurate responses (i.e. there is less wandering).

\section{\label{sec:discussion}Discussion}
Metastability is a powerful mechanism for supporting robust representation of information in neural circuits~\cite{la2019cortical,brinkman2022metastable,bressloff2010metastable}. Recent works have begun to explore possible roles of multistability in neuron firing arising from cell mechanisms such as nonlinear denditric compartmentalization ~\cite{goldman2003robust} and intrinsic subthreshold oscillation bands ~\cite{champion2023oscillatory} which link step-like firing rate sets of neurons and graded amplitudes of neuron activity. On the macroscale, we have proposed a neural circuit model inspired by microclustered architecture~\cite{bressloff2010metastable,litwin2012slow} of such multistable neurons with population firing dependent on cluster activation~\cite{mazzucato2015dynamics} which sustains neural population activity bumps with multiple amplitudes. Rather than employing a fragile model with a fine-tuned transfer function~\cite{carroll2014encoding,esnaola2022flexible,kutschireiter2023bayesian}, we considered a quantized firing rate function generating a robust model reminiscent of metastable single neuron models with bistable dendritic compartments~\cite{goldman2003robust}. Our neural circuit model's dynamics can be reduced to evolution equations that clearly account for how stochasticity and perturbations impact delay encoding and confidence. Our analysis provides a simple and understandable theory for increased accuracy of delayed estimates made from more salient cues~\cite{basso1997modulation,badcock2008examining,cohen2014correspondence}. 

Our neural field model can support up to $N$ pairs of stable/unstable bumps when its stairstep firing rate function possesses $N$ steps. Active solutions of the neural field asymptotically relax to similarly shaped bump profiles, strongly suggesting an ansatz for low-dimensionalizing system dynamics. Both external inputs and noise can drive neural activity bumps between neighboring amplitude values, as described by our reduced system. Our reduced equations not only accurately predict the wandering of bumps in response to noise, but the timing and preference of amplitude transitions. We find bumps formed from more salient cues are more resilient to fluctuations and generate more accurate response estimates.

Our analysis could be extended in several ways. Short term plasticity can further stabilize bumps during delay periods~\cite{Kilpatrick18STP_FacSCIREP}, effects that could be analyzed using an interface based analysis~\cite{cihak2024multiscale}. We could also consider models with separate excitatory and inhibitory populations and develop theory separately tracking each bump's phase and amplitude dynamics~\cite{cihak2022distinct}. Our phase-amplitude ansatz makes near-equilibrium assumptions about the shape of the bump, but perturbations may warp the bump profile in ways not well described by multiplicative scalings. Consideration of such additive changes to neural population responses could more fully characterize the continuum of spontaneous modulations to neural tuning~\cite{arandia2016multiplicative}. Accounting for dynamic perturbations of bump profiles may also improve the accuracy of our amplitude transition theory.
Also, our analysis is limited to firing rate functions with a few steps, $N = \mathcal{O}(1)$, but could be extended to examine the limit of many steps ($N \gg 1$) with frequent transitions between or the impact of more irregular step spacings.

Quantized representation of inputs and even behavior variables is supported by a number of other computational and experimental studies. Clustered spiking networks can also generate staircase firing rate relations~\cite{litwin2012slow,mazzucato2015dynamics}, suggesting that a mean field analysis of such models could be a fruitful direction of future study. Complementarily, modular cortical networks exhibiting clustering at larger scales have also demonstrated improved working memory performance~\cite{Curtis06, constantinidis2004neural, kilpatrick2015delay,Kilpatrick13FrontiersMultiArea}, further supporting the hypothesis that architectures that engender metastability could underlie more robust delay encoding. More recently, an analysis of neural activity in hypothalamus has revealed that general behavioral states like aggression may be encoded along a discrete and approximate line attractor~\cite{nair2023approximate}. It is important to note that the destruction of a line attractor may not be catastrophic for coding delayed estimates or other states as long as the resulting evolution of the dynamics is slow compared to the needed encoding time~\cite{sussillo2013opening}.
% Such a theory is supported by multilayer neural field models with redundant bump encodings~\cite{kilpatrick2015delay,Kilpatrick13FrontiersMultiArea}. 

% Complimentarily, \cite{tang2019working} observed that prefrontal neurons become more active and less variable as working memory performance improved over training, indicating distributed activation and reduced variability play a role in the neural basis of working memory.  

Our neuromechanistic model provides several links between circuit features and behavioral response trends, providing a testbed for physiological theories of increased errors and impaired processing for continuum estimates in schizophrenia~\cite{badcock2008examining, stein2021towards}, autism spectrum disorders~\cite{vattikuti2010computational}, or attention deficit hyperactivity disorder~\cite{castellanos2000executive}. Early detection of such abnormalities using non-invasive psychophysics could speed diagnoses and the implementations of behavioral interventions to help manage executive function in neurodivergent populations~\cite{klingberg2010training,kluwe2013executive}. Models that can connect aberrant response statistics to underlying neurophysiology require carefully balancing mathematical tractabilty and the inclusion of hitherto unexplored features of the underlying biological circuits.

%\bibliography{ref}	
%\input{metastable_bib.tex}

\providecommand{\noopsort}[1]{}\providecommand{\singleletter}[1]{#1}%

\appendix

\section{Model equation simulations}\label{app:sims}
Python code for simulating and analyzing our neural field models are available at the repository: \url{https://github.com/cihakh/RobustCertaintyEncoding}.

Convolutions and spatially filtered noise were computed using fast Fourier transforms. Euler-Maruyama is used to time-step Eq.~(\ref{eq:model_C4}) with initial conditions and inputs centering bumps at $x=0$. Numerical quadrature is performed using Riemann sums. Table \ref{tableparam_C4} gives simulation parameters unless otherwise indicated in figure captions.

Amplitude transition times are found by (1) initializing simulations starting with a bump having an amplitude corresponding to a stable stationary bump; (2) running a stochastic simulation until the estimated amplitude crosses through a neighboring unstable bump value (or until a maximum time is reached); (3) recording the time of transition for 1000 trials with transitions detected within the delay or terminating when 25 successive or 100 cumulative trials have failed to transition, which we take to indicate that the mean transition time is too close to or far beyond the cutoff time for our parameterized method to make an accurate estimate. Bump amplitudes are estimated as the peak activity value $\max_xu(x,t)$.

\begin{figure}[b!]
    \centering
    \includegraphics[width=0.45\textwidth]{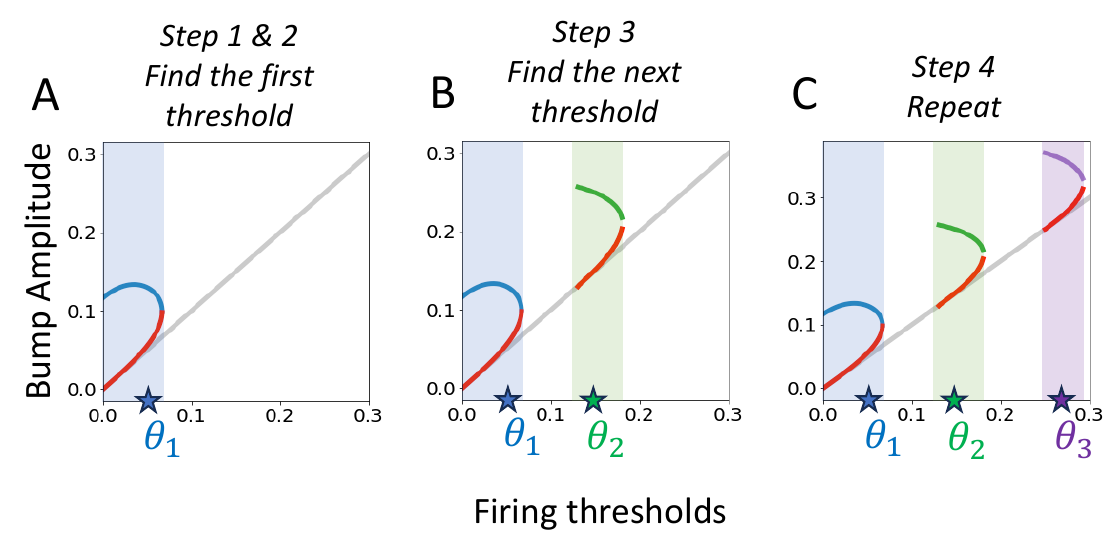}
    \caption[Choosing stairstep firing thresholds]{\textbf{Iteratively identifying bumps and stairstep firing thresholds.} We briefly illustrate the sequential procedure of determining stable/unstable bump branches and stairstep firing thresholds. \textbf{A.}~Select $\theta_1$ and find the $B=1$ set of bump solution branches (blue branch is stable); \textbf{B.}~After choosing $\theta_1$ find the second set of branches (green/red lines in green region). \textbf{C.}~Iterate for thresholds $\theta_{3:N}$. Gaps between solution regions arise since $B=k+1$ bumps must all have higher amplitudes than $B=k$ bumps (lowest point on neighboring red branch sits above blue branch).}
    \label{fig:iterthresh}
\end{figure}

\begin{figure*}[t!]
    \centering
    \includegraphics[width=0.75\textwidth]{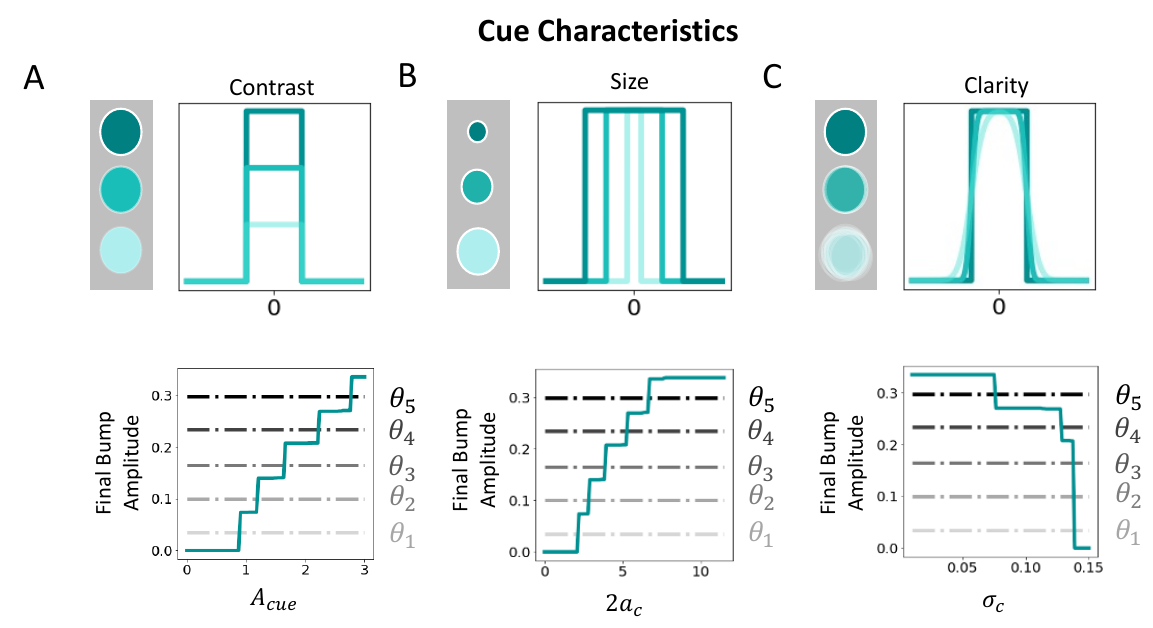}
    \caption[Cue characteristics and certainty]{\textbf{Cue characteristics and certainty.} Cue profile ($A_c(t)\left[ {\rm erf} \frac{x+a_c}{\sigma_c} - {\rm erf} \frac{x-a_c} {\sigma_c} \right]$ portion of $I_c(x,t)$) features are varied and resulting bump amplitude determined. \textbf{A.}~Contrast controls cue strength $A_{cue}$ (bottom panel), whose increase generates higher amplitude bumps. \textbf{B.}~Wider cues generated by increasing the diameter $2a_c$ increase bump amplitude. \textbf{C.}~Cue clarity (sharpness) is reduced by decreasing $\sigma_c$ in Eq.~(\ref{eq:cue}), decreasing encoded bump amplitude.}
    \label{fig:features_vs_amplitudes}
\end{figure*}

\section{Iterative construction of bumps} \label{iterexist}
Bumps are constructed by identifying threshold intervals in which solutions of successively higher amplitude exist. Starting by solving the threshold condition Eq.~(\ref{threshcond}) at the first level ($i=1$) and constraining $\theta_1 < U(0)$, we can find $\theta_1$ values that admit stable/unstable branches of $B=1$ bumps. Then, finding the peak of the maximum $U(0)$ for the stable $B=1$ bumps, we constrain an interval of possible $\theta_2$ values and use the Fourier coefficient equations Eq.~(\ref{Uj}) to determine the next larger family of bumps of sufficiently high amplitude so that $\theta_2 < U(0)$. For a satisfactory $\theta_2$, we can continue branches of stable/unstable bumps. This process is then repeated by choosing an appropriate $\theta_{k+1}> U(0)$ for all $B=k$ bumps calculated from the Fourier decomposition given by Eq.~(\ref{Uj}), using Eq.~(\ref{Uj}) to compute the next branches such that $U(0) > \theta_{k+1}$ until $k+1=N$. See Fig.~\ref{fig:iterthresh} for an illustration.

\section{Amplitude dependence on cue characteristics}\label{app:cue_char} 
Here we qualitatively compare bump amplitudes to other cue properties. Experiments show longer encoding periods lead to higher accuracy and increased neural responses~\cite{badcock2008examining, groen2022temporal},  consistent with longer cue durations generating bumps of higher amplitude (Fig.~\ref{fig:coeff_ampl_evolution}A). 
Increasing cue contrast (stimulus amplitude) also increases neural responses (See \cite{groen2022temporal, mcadams1999effects} and Fig.~\ref{fig:features_vs_amplitudes}A). Larger cues can also elicit higher neural responses (See \cite{dobbins1998distance} and Fig.~\ref{fig:features_vs_amplitudes}B). Cue blurriness can also impact detection and encoding (See \cite{mcgettigan2012speech,cunningham2002effects} and Fig.~\ref{fig:features_vs_amplitudes}C).

\end{document}